\documentclass[amsmath,amssymb,prc,showpacs]{revtex4}

\usepackage{bm,hyphenat,xspace}
\usepackage{graphicx,epsfig}
\usepackage{soul}

\newcommand {\mcu}{\mathcal{U}}

\newcommand{\He}{{}^3\mathrm{He}}

\newcommand{\nH}{n\text{-}{}^3\mathrm{H}}
\newcommand{\pHe}{p\text{-}{}^3\mathrm{He}}
\newcommand{\pH}{p\text{-}{}^3\mathrm{H}}
\newcommand{\nHe}{n\text{-}{}^3\mathrm{He}}

\newcommand{\dd}{d\text{-}d}
\newcommand{\AB}{A\text{-}B}

\newcommand{\nuc}[2]{\ensuremath{\rm{^{#1}}#2}}  
\begin{document}

\title {Benchmark calculation of $\pH$ and $\nHe$ scattering}

\author{M.\ Viviani$^{\,{\rm a}}$,
        A. Deltuva$^{\,{\rm b}}$,
        R. Lazauskas$^{\,{\rm c}}$,
        A.~C.~Fonseca$^{\,{\rm d}}$,
        A.\ Kievsky$^{\,{\rm a}}$,
  and L.E.\ Marcucci$^{\,{\rm  a,f}}$}

\affiliation{
$^{\,{\rm a}}$ INFN-Pisa, 56127 Pisa, Italy\\
$^{\,{\rm b}}$ Institute of Theoretical Physics and Astronomy, Vilnius
  University, LT-10222 Vilnius, Lithuania\\
$^{\,{\rm  c}}$ IPHC, IN2P3-CNRS/Universit\'e Louis Pasteur BP 28,
              F-67037 Strasbourg Cedex 2, France\\
$^{\,{\rm d}}$  Centro de F\'isica Nuclear da Universidade de Lisboa,
              P-1649-003 Lisboa, Portugal\\
$^{\,{\rm f}}$ Department of Physics, University of Pisa, 56127 Pisa, Italy
}

\received{\today}

\begin{abstract}
$\pH$ and $\nHe$  scattering in the energy range above the $\nHe$ but below
the $d-d$ thresholds is studied by solving the
4-nucleon problem with a realistic nucleon-nucleon interaction. Three different
methods -- Alt, Grassberger and Sandhas, Hyperspherical
Harmonics, and Faddeev-Yakubovsky -- have been employed
and their results for
both elastic and charge-exchange processes are compared. We
observe a good agreement between the three different methods, thus
the obtained results may serve as a benchmark. A comparison with the
available experimental data is also reported and discussed.
\end{abstract}

 \maketitle


\section{Introduction}
\label{sec:intro}

Modern studies of nuclear structure and dynamics are mostly based
on {\it ab initio} calculations using realistic potentials. Due to
the complexity of the problem, it is clearly important to have
benchmarks between different groups and different techniques in
order to test validity of the existent codes, as well as to
establish the numerical accuracy of the solutions. On its turn
this may allow a meaningful comparison with experimental data and
then serve as a probe on our current understanding of the nuclear
dynamics. In particular, this program is well suited to be pursued
in few-nucleon systems ($A\le6$), where several well controlled
numerical techniques have been developed.

The interest in {\it ab initio} calculations has been renewed in
recent years, i.e. after the advent of the theoretical framework
of chiral effective field theory ($\chi$EFT), nowadays widely used
to derive nuclear potentials and electroweak currents from the
symmetries of QCD---the exact Lorentz, parity, and time-reversal
symmetries, and the approximate chiral symmetry~(see, for example,
Refs.~\cite{Weinberg90,Ordonez96,Epelbaum09,ME11}). The test of
these new potentials in few-nucleon scattering, where accurate
measurements of several observables exist, will give very
stringent and critical information.

The three-nucleon system is thoroughly studied and for this case
some very accurate benchmarks~\cite{Deltuva05,Witala06} exist.
After this achievement focus has been set on the four-nucleon (4N)
sector. In first place, this system may serve as an ideal
``theoretical laboratory" to test our knowledge of the
nucleon--nucleon (NN) and three--nucleon (3N) interactions. In
particular, the effects of the NN P-waves and of the 3N force are
larger than in the $A=2$ or $3$ systems, and it is the simplest
system where the 3N interaction in channels of total isospin
$T=3/2$ can be studied. In second place, there is a number of
reactions involving four nucleons which are of extreme importance
for astrophysics, energy production, and studies of fundamental
symmetries. As an example, the reactions $n + \nuc{3}{He}$
and $d + d $ play a key role in the theory of big-bang
nucleosynthesis.

Moreover, the potentials derived from $\chi$EFT contain several
unknown parameters -- the so-called low energy constants (LECs) --
which have to be fixed by comparison with experimental data. This
fitting procedure is usually brought forth in $A=2$ and $3$
systems, where accurate calculations can be performed since many
years and abundant precise experimental data exist. It is
therefore of great interest to test validity of these $\chi$EFT
potentials on independent data, which has not been included in the
parametrization. The 4N system, containing several resonances
which are not straightforwardly correlated with NN and 3N sector,
presents an ideal testground.

Nowadays, the 4N bound state problem can be numerically solved with
good accuracy. For example, in Ref.~\cite{Kea01} the binding energies and
other properties of the $\alpha$-particle were studied using the
AV8$'$~\cite{AV18+} NN interaction; several different techniques produced
results in very close agreement with each other (at the level of less than
1\%). More recently, the same agreement has also been obtained  considering
different realistic NN+3N force models~\cite{Wea00,Nogga03,Lazaus04,Viv05}.

In recent years, there has also been a rapid advance in solving the 4N
scattering problem with realistic Hamiltonians. Accurate calculations of four-body
scattering observables have been achieved in the framework of the
Alt-Sandhas-Grassberger (AGS) equations~\cite{grassberger:67}, solved in momentum
space \cite{DF07,DF07b,DF07c}, where the long-range Coulomb interaction is treated using the
screening and renormalization method \cite{Alt78,DFS05}.
Also solutions of the Faddeev-Yakubovsky (FY) equations in
configuration space~\cite{Cie98,Rimas_03,LC_FB17_03,Lea05,Lazaus09} and the application of the
Hyperspherical Harmonics (HH) expansion method~\cite{rep08} to the solution
of this problem have been reported~\cite{Vea09,Vea13}.
In addition to these methods, the solution of the 4N scattering
problem has been obtained also by using techniques based on the
resonating group model~\cite{HH97,PHH01,HH03,Sofia08}. For
applications to $A>4$ systems, see Ref.~\cite{Petr16} and references
therein.

In a previous work, we have presented a benchmark calculation for
low energy $\nH$ and $\pHe$ elastic observables by using the
aforementioned AGS, FY, and HH techniques, and by employing
different NN interactions~\cite{Deltuva11}. Nice agreement between
the results of the three different calculations has been reported,
only minor differences were observed for some small polarization
observables. It has been concluded that the $\nH$ and $\pHe$
elastic scattering problem can be nowadays solved with a good
accuracy.

In the present paper, we extend the benchmark to $\pH$ and $\nHe$
scattering for energies where both channels are open but below the
$d-d$ threshold. These calculations present new
challenges and are rather complex since the two reaction channels
are coupled and involve both total isospin $T=0$ and $T=1$ states.
So far, only a few accurate calculations have been performed for
these processes~\cite{DF07c,Lazaus09}. Only recently, the AGS method 
has been extended to the energy regime well above
breakup threshold where the calculations become even more complicated
due to nontrivial boundary conditions or singularities
\cite{Deltuva15}.  Therefore, we
consider the present benchmark as an important step in
establishing our current capability to solve the $A=4$ scattering
problem. Moreover, our aim is also to provide a set of solid
converged results which could represent useful benchmarks for
future applications in $A=4$ scattering. The potential used in
this paper is the N3LO500 model by Entem and
Machleidt~\cite{EM03}, based on the $\chi$EFT approach and derived
 up to next-to-next-to-next-to-leading
order in chiral perturbation theory.

In addition to the desire of improving theoretical tools, it is
important to note that for these reactions there exist a large
amount of accurate experimental data, accumulated during the last
50 years. In particular, below the $d-d$ threshold there exist
(since many years) several measurements of the $\pH$ elastic
differential cross
section~\cite{Hemme49,Clas51,Jarmie59,Brol64,Mandu68,Iva68,Kanko76},
$\nHe$ elastic differential cross section~\cite{Sea60,Say61} and
total cross section~\cite{Sea60,Say61,Alfi81,Haes83}, and 
$\nHe$ elastic analyzing powers~\cite{Holla72,Sinra76,Klage85,Jany88,Este13}.
Regarding the $\nHe \rightarrow  \pH$ charge exchange reaction,
there exist mainly measurements of the total cross
section~\cite{Coon50,Batch55,Gibb59,Say61,Als64,Mack65,Cost70,Borza82,Haes83}.
For the $\pH \rightarrow  \nHe$ charge exchange reactions, there
exist measurements of the differential cross
section~\cite{Jarvis50,Will53,Jarvis56,Drosg80} and polarization
observables~\cite{Cra71,Doy81,Tornow81,Walst98,Wilburn98}.
Therefore, another motivation of the present work is to compare
theoretical predictions with this data.

The paper is organized as follows. In Section~\ref{sec:methods},
a brief description of the methods used for this benchmark is
reported. {\ In Section~\ref{sec:results}, a comparison  between
the results obtained within the different schemes is shown
and the theoretical calculations are also compared
with the available experimental data. The conclusions will be
provided in Section~\ref{sec:conc}. }

\section{Methods}
\label{sec:methods} In this work three different techniques,
namely the AGS equations, the HH method and the FY equations, will
be employed to solve 4N scattering problem and the results
provided by the three approaches will be benchmarked. Generalities
common to the three methods 
will be discussed in this section, whereas technicalities proper
to each technique will be presented separately in devoted
sections.

In case of a two body clustering $\AB$, the total energy of the
scattering state in the center-of-mass (CM) system is given by
\begin{equation}
  E=-B_A-B_B+T_{CM}\, \label{eq:energy}
\end{equation}
where
\begin{equation}
  T_{CM}={q_\gamma^2\over 2\mu_\gamma}\ , \qquad
  {1\over \mu_\gamma} = {1\over M_A}+{1\over M_B}
   \ ,\label{eq:tcm}
\end{equation}
$q_\gamma$ is the relative momentum between clusters,
and $M_X$ ($B_X$) is the mass (binding energy) of the cluster $X$.
Clearly, in the case of a single nucleon $M_X=M_N$, where $M_N$ is
the nucleon mass, and $B_X=0$.  In this paper, we limit ourselves
to study the scattering for $-B_{\He}<E<-2B_d$, where $B_d$ is the
deuteron binding energy. Namely, we consider 4N scattering when
the channels $\pH$ and $\nHe$ are open, but the  $\dd$ channel is
closed.

In the following, $\gamma$ will denote the particular asymptotic
clustering $\AB$. More specifically, $\gamma=1$ $(2)$ will
correspond to the $\pH$ ($\nHe$) asymptotic clustering. Moreover,
for example when discussing $\nHe$ scattering, the observables
will be calculated at a given neutron laboratory energy $E_n$,
corresponding approximately to $E_n\approx (4/3) T_{CM}$.

For a given total angular momentum quantum number $J$ and parity
  $\pi$, the information on the scattering observables is contained in the
  S-matrix  ${\cal S}^{J\pi}_{\gamma LS,\gamma^\prime L^\prime
      S^\prime}$, where $\gamma L S$ ($\gamma^\prime L^\prime
    S^\prime$) denote the initial (final) clustering type,
    relative orbital momentum and channel spin of the two clusters,
    respectively (see below).
Each submatrix of the S-matrix representing a separate cluster is
clearly no longer unitary. For example, the submatrix describing
$\nHe$ elastic scattering will be denoted as ${\cal
S}^{J\pi}_{\gamma=2\; LS,\gamma'=2\;L^\prime S^\prime}(E)\equiv
{\cal S}^{\nHe,J\pi}_{LS,L^\prime S^\prime}(E)$. For the
$J^\pi=0^\pm$ channels, submatrix  ${\cal S}^{\nHe,J\pi}$ is of
dimension 1 and can be parametrized in the standard way
\begin{equation}
  {\cal S}^{\nHe,J\pi}_{LS,LS} = \eta^{J\pi}_{LS} \exp(2i \delta^{J\pi}_{LS}) \ .
  \label{eq:sma1}
\end{equation}
For the other cases, ${\cal S}^{\nHe,J\pi}$ is of dimension 2 and
is conveniently parametrized as
\begin{equation}
  {\cal S}^{\nHe,J\pi}=
 \left(\begin{array}{cc}
     a & b+ic \\
     -(b+ic) & a \\
   \end{array}\right)^{-1}
 \left(\begin{array}{cc}
     \eta_{LS}^{J\pi} \exp(2i \delta_{LS}^{J\pi}) & 0 \\
      0 &  \eta_{L'S'}^{J\pi} \exp(2i \delta_{L'S'}^{J\pi}) \\
   \end{array}\right)
  \left(\begin{array}{cc}
     a & b+ic \\
     -(b+ic) & a \\
   \end{array}\right) \ ,
\end{equation}
where the (eigen)phase-shifts $\delta_{LS}^{J\pi}$, the
(eigen)inelasticity parameters $ \eta_{LS}^{J\pi}$, and the
parameters $a$, $b$, and $c$ are real (and $a^2+b^2+c^2=1$).
Explicitly, $(a,b+ic)$ and $(-b-ic,a)$ are the (right)
eigenvectors of the matrix $ {\cal S}^{\nHe,J\pi}$ associated to
the two eigenvalues, the latter written as $\eta_{LS}^{J\pi}
\exp(2i \delta_{LS}^{J\pi})$ and $\eta_{L'S'}^{J\pi} \exp(2i
\delta_{L'S'}^{J\pi})$. Then, we parametrize
$a+i(b+ic)\equiv\exp(i\epsilon^{J\pi})$. If the matrix $ {\cal
  S}^{\nHe,J\pi}$ would be unitary, one recovers the standard
definition of mixing parameter $a=\cos(\epsilon^{J\pi})$,
$b=\sin(\epsilon^{J\pi})$, $c=0$ with $\epsilon^{J\pi}$ real.
On the other hand, with inelastic channels present, $\epsilon^{J\pi}$
is complex.

\subsection{AGS Equations}
\label{sec:AGS}

The AGS equations \cite{grassberger:67}
for the four-body transition operators were derived  assuming
short-range interactions, but together with the
screening and renormalization method  \cite{Alt78,DFS05,DF07b},
they can be applied also to systems with repulsive Coulomb force.
The isospin formalism enables the symmetrization of
the AGS equations \cite{DF07} in the 4N system, where there are only two
distinct four-particle partitions, one of the $3+1$ type and
one of the $2+2$ type, denoted  by $\alpha =1$ and $2$, respectively.
 In terms of particles 1,2,3,4 we choose those partitions to be (12,3)4 and (12)(34),
respectively. The corresponding transition operators $\mcu_{\beta\alpha}$
for the initial states of the $3+1$ type, as appropriate
for the $\nHe$ and $\pH$ scattering, obey the integral equations
\begin{eqnarray}\label{eq:Uags}
\mathcal{U}_{11}  &= {}& -(G_0 \, T  G_0)^{-1}  P_{34}
 - P_{34} \, U_1\, G_0 \, T G_0 \, \mathcal{U}_{11} 
 + {U}_2   G_0 \, T G_0 \, \mathcal{U}_{21} ,
\label{eq:U11} \\
\mathcal{U}_{21}  &= {}&  (G_0 \, T  G_0)^{-1} \, (1 - P_{34}) 
+ (1 - P_{34}) U_1\, G_0 \, T  G_0 \, \mathcal{U}_{11} .  \label{eq:U21}
\end{eqnarray}
Here $G_0 = (E+i0 -H_0)^{-1}$ is the free resolvent, $H_0$ is the free Hamiltonian,
$P_{ij}$ is the permutation operator of particles $i$ and $j$,
 $T=V+V G_0T$ is the two-nucleon transition matrix derived from the potential $V$, and
\begin{equation}
\label{eq:U}
U_{\alpha}  = {}  P_\alpha G_0^{-1} + P_\alpha \, T G_0 \, U_{\alpha}, \\
\end{equation}
are $3+1$ and $2+2$ subsystem transition operators with
$P_1  = {}  P_{12}\, P_{23} + P_{13}\, P_{23}$ and
$P_2  = {}  P_{13}\, P_{24}$.

The integral AGS equations (\ref{eq:Uags}) are solved in momentum-space
partial-wave framework.
Scattering amplitudes for elastic and charge-exchange reactions
are given by on-shell matrix elements of $ \mathcal{U}_{11}$
as described in Refs. \cite{DF07,DF07c,Deltuva15}
where also further details regarding the numerical solution
can be found. Note that in the considered energy regime
the only singularities in the kernel of AGS equations
arise due to bound-state poles of $U_1$ and are treated
by a simple subtraction method.


\subsection{HH Expansion}
\label{sec:HH}

The wave function describing a scattering process with incoming clusters
specified by the index $\gamma$ and in a state of total angular momentum quantum numbers $J,J_z$,
relative orbital angular momentum $L$, and channel spin $S$
($S=0, 1$) can be written as
\begin{equation}
    \Psi_{1+3}^{\gamma LS,JJ_z}=\Psi_C^{\gamma LS,JJ_z}+\Psi_A^{\gamma LS,JJ_z} \ ,
    \label{eq:psica}
\end{equation}
keeping in mind the  notation $\gamma=1$ $(2)$  to represent
respectively $\pH$ ($\nHe$) asymptotic clustering.

The ``core'' part of wave function $\Psi_C^{\gamma LS,JJ_z}$
describes the system in the region where four particles are close
to each other and where their mutual interactions are strong.
Hence, $\Psi_C^{\gamma LS,JJ_z}$ vanishes in the limit of large
inter-cluster distances. This part of the wave function is written
as a linear expansion $\sum_\mu c^{\gamma LSJ}_\mu {\cal Y}_\mu$,
where ${\cal Y}_\mu$ is a set of basis functions constructed in
terms of the HH functions (for more details see, for example,
Ref.~\cite{rep08}).

The other part $\Psi_A^{\gamma LS,JJ_z}$ describes the relative
motion of the clusters in the asymptotic regions, where
these clusters do not interact (except eventually for the
long-range Coulomb interaction). In the asymptotic region the wave
function $\Psi_{1+3}^{\gamma LS,JJ_z}$ reduces to
$\Psi_{A}^{\gamma LS,JJ_z}$, which must therefore be the
appropriate asymptotic solution of the Schr\"odinger equation.
Then, $\Psi_{A}^{\gamma LS,JJ_z}$ can be decomposed into a linear
combination of the following functions
\begin{eqnarray}
  \Omega_{\gamma LS}^\pm &=& {\cal A}\biggl\{
  \Bigl [ Y_{L}(\hat{\bm y}_\gamma) \otimes  [ \phi_A \otimes \phi_B]_{S}
   \Bigr ]_{JJ_z} \nonumber\\
  &&\times \left ( f_L(y_l) {\frac{G_{L}(\eta_\gamma,q_\gamma y_\gamma)}{q_\gamma y_\gamma}
          \pm {\rm i} {\frac{F_L(\eta_\gamma,q_\gamma y_\gamma)}{q_\gamma y_\gamma}}} \right )\biggr\} \ ,
  \label{eq:psiom}
\end{eqnarray}
where  $y_\gamma$ is the distance between the CM
of clusters $A$ and $B$, $q_\gamma$ is the magnitude of the
relative momentum between the two clusters (see
Eq.~(\ref{eq:tcm})), and $\phi_A$ and $\phi_B$ the corresponding
bound state wave functions. In the present work, the trinucleon
bound state wave functions are calculated very accurately by means
of the HH method~\cite{Nogga03,rep08} using the corresponding
$A=3$ Hamiltonian. Conventionally, we identify the trinucleon
bound state wave function with $\phi_A$. Therefore,  $\phi_B$ describes
the single nucleon spin-isospin state. The channel spin $S$ is obtained by coupling
the angular momentum of the two clusters. In our case, clearly
$S=0,1$. The symbol ${\cal A}$ means that the expression between
the curly braces has to be properly antisymmetrized.

In Eq.~(\ref{eq:psiom}), the functions $F_L$ and
$G_{L}$ describe the asymptotic radial motion of the
clusters $A$ and $B$. If the two clusters are composed of $Z_A$ and $Z_B$
protons, respectively, the parameter $\eta_\gamma=\mu_\gamma Z_A Z_B
e^2/q_\gamma$, where $e^2\approx 1.44$ MeV fm. The function $F_L(\eta,qy)$ is
the regular Coulomb function while $G_{L}(\eta,qy)$ is
the irregular Coulomb function. The function $f_L(y)=[1-\exp(-\beta
  y)]^{2 L+1}$ in Eq.~(\ref{eq:psiom})
has been introduced to regularize  $G_L$  at small $y$, and
$f_L(y) \rightarrow 1$ as $y$ is large, thus  not affecting the asymptotic
behavior of $\Psi_{1+3}^{\gamma LS,JJ_z}$. Note that for large values of $qy_l$,
\begin{eqnarray}
  \lefteqn{ f_L(y_l) G_{L}(\eta,qy_l)\pm {\rm i} F_L(\eta,qy_l) \rightarrow
   \qquad\qquad} &&  \nonumber \\
  && \exp\Bigl[\pm {\rm i} \bigl (q y_l-L\pi/2-\eta\ln(2qy_l)+\sigma_L\bigr )\bigr ] \ ,
\end{eqnarray}
where $\sigma_L$ is the Coulomb phase-shift. If $\eta$ is zero,
the Coulomb functions reduce to the Riccati-Bessel
functions~\cite{abra}. Therefore, $\Omega_{\gamma LS,JJ_z}^+$
($\Omega_{\gamma LS,JJ_z}^-$) describe the asymptotic outgoing
(incoming) $A-B$ relative motion. Finally,  $\Psi_A^{\gamma
    LS,JJ_z}$ is given by
\begin{equation}
  \Psi_A^{\gamma LS,JJ_z}= \Omega_{\gamma LS,JJ_z}^- -
     \sum_{\gamma^\prime L^\prime S^\prime}
     {\cal S}^{J\pi}_{\gamma LS,\gamma^\prime L^\prime S^\prime}(E)\;
     \Omega_{\gamma^\prime L^\prime S^\prime,JJ_z }^+  \ ,
  \label{eq:psia}
\end{equation}
where the parameters ${\cal S}^{J\pi}_{\gamma LS,\gamma^\prime
L^\prime S^\prime}(E)$ are the $S$-matrix elements at the energy
$E$, given by Eq.~(\ref{eq:energy}).  Of course, the sum over
$L^\prime$ and $S^\prime$ is over  all values compatible with the
given $J$ and parity $\pi$. In particular, the sum over $L^\prime$
is limited to include either even or odd values such that
$(-1)^{L^\prime}=(-1)^L=\pi$. The sum over $\gamma^\prime$ is
limited to the {\it open} channels (namely those channels for
which $q_\gamma^2>0$, see Eq.~(\ref{eq:tcm})).  For the scattering
process considered in the present paper, clearly $\gamma'=1$, $2$.

The $S$-matrix elements ${\cal S}^{J\pi}_{\gamma
LS,\gamma'L^\prime S^\prime}(E)$ and coefficients $c^{\gamma
LSJ}_\mu$ occurring in the HH expansion of $\Psi^{\gamma
LS,JJ_z}_C$ are determined by forming a functional
\begin{equation}
   [{\cal S}^{J\pi}_{\gamma LS,\gamma^\prime L^\prime S^\prime}(E)]=
    {\cal S}^{J\pi}_{\gamma LS,\gamma^\prime L^\prime S^\prime}(E)
     -{1\over 2i}
        \left \langle\Psi^{\gamma^\prime L^\prime S^\prime,JJ_z}_{1+3} \left |
         H-E \right |
        \Psi^{\gamma LS,JJ_z}_{1+3}\right \rangle
\label{eq:kohn}
\end{equation}
stationary with respect to variations in ${\cal S}^{J\pi}_{\gamma
  LS,\gamma^\prime L^\prime S^\prime}$ and $c^{\gamma LSJ}_{\mu}$ (Kohn variational principle).
By applying this principle, a linear set of equations
is obtained for ${\cal S}^{J\pi}_{\gamma LS,\gamma^\prime L^\prime S^\prime}$ and
$c^{\gamma LSJ}_{\mu}$. This linear system is solved using the Lanczos
algorithm.

This method can be applied in either coordinate or momentum
space, and using either local or non-local
potentials~\cite{rep08}. The
first steps are (1) the use of the method discussed in
Ref.~\cite{V98} to antisymmetrize the HH functions and (2) 
a partial wave decomposition of the asymptotic functions
$\Omega_{\gamma LS,JJ_z}^\pm$, the latter task being rather time
consuming. After this decomposition, the calculation of the matrix
element in Eq.~(\ref{eq:kohn}) is fast, except for the $J^\pi=2^-$
state, due to the large number of HH functions to be included in the
expansion in this particular case. After these steps,
the problem reduces to the solution of a linear system.

The expansion of the scattering wave function in terms of the HH basis
is in principle infinite, therefore a truncation scheme is necessary.
The HH functions are essentially characterized by
the orbital angular momentum
quantum numbers $\ell_i$, $i=1,3$, associated with the three Jacobi
vectors, and the grand angular quantum number $K$ (each HH
function is a polynomial of degree $K$). The basis is
truncated to include states with $\ell_1+\ell_2+\ell_3\le \ell_{\rm
max}$ (with all possible re-coupling between angular and spin states appropriate to
the given $J$). Between these states, we retain only the HH functions with
$K\le K_{\rm max}$. Note that in the calculation we have included
both states with total isospin $T=0$ and $1$.

The main sources of 
numerical uncertainties for this method could come from  the numerical integration needed
to compute the matrix elements of the Hamiltonian and the truncation
of the basis. It has been checked that the
numerical uncertainty of the calculated phase-shifts related to the numerical
integration is small (around $0.1$ \%). The NN interaction has been
limited to act on two-body states with total angular momentum $j\le
j_{\rm max}=8$ (greater values of $j_{\rm max}$ are completely
negligible).  The largest uncertainty is related to the use of a
finite basis due to the slow convergence of the results with
$K_{\rm max}$.  This problem can be partially overcome by performing
calculations for increasing values of $K_{\rm max}$ and then using some
extrapolation rule (see for example Ref.~\cite{Fisher06}) to get the
``$K_{\rm max}\rightarrow\infty$'' result. This procedure 
introduces a new  uncertainty which can be
estimated. A detailed study of this problem will be published
elsewhere~\cite{Vea16}. The convergence of the quantities of interest
in term of $K_{\rm max}$ is slower using NN potentials with a strong
repulsion at short interparticle distance, but it
is less relevant for the N3LO500 potential. For the present case,
this uncertainty has been estimated to be at most $0.5$ \%.

The convergence with $\ell_{\rm max}$ is usually rather fast and
values of $\ell_{\rm max}$ around either $5$ or $6$ have been found to be
sufficient.  However,  in some cases, we have found a slow
convergence  of the inelasticity parameters $\eta_{LS}^{J\pi}$. To
give an example, in Table~\ref{tab:eta11}, we report the values of
the $\nHe$ $\eta_{LS}^{J\pi}$ and $\delta_{LS}^{J\pi}$
parameters for the $J^\pi=0^-$
wave, calculated  with the HH method as a function of $\ell_{\rm
max}$. The calculations have been performed at $E_n=1$ MeV and for
the N3LO500 potential. For this wave the parity is negative, so
only HH functions having $\ell_{\rm max}$ odd have to be included
in the expansion. As can be seen, the inclusion in the expansion
of the ``core'' part of functions with $\ell_{\rm max}=1$ is
insufficient to obtain a reasonable estimate of these parameters.
The addition of also the $\ell_{\rm max}=3$ functions improves
noticeably the calculation of the phase-shift  $\delta_{11}^{0-}$,
which is now very close to the final result (in the table, we have
also reported the same parameters obtained using the AGS
equations). The inclusion of the $\ell_{\rm max}=5$ brings finally
the HH results in agreement with that obtained by the AGS method.
On the other hand, the $\eta_{11}^{0-}$ values appears to converge
slowly. Note that for $\pHe$ scattering (see our previous
benchmark~\cite{Deltuva11}) only one asymptotic channel is open,
so in all calculations one obtains $\eta_{11}^{0-}=1$ with a very
good accuracy.

\begin{table*}[t]
\begin{ruledtabular}
\begin{tabular}{*{3}{r}}
 $\ell_{\rm max}$ & $\eta_{11}^{0-}$  & $\delta_{11}^{0-}$ (deg) \\
\hline
 1 & 0.603 & +21.9 \\
 3 & 0.459 & -\phantom{0}9.9 \\
 5 & 0.505 & -10.5 \\
\hline
AGS & 0.553 & -10.5 \\
\end{tabular}
\end{ruledtabular}
\caption{ \label{tab:eta11} $\nHe$ inelasticity parameter
$\eta_{11}^{0-}$ and  phase-shift
  $\delta_{11}^{0-}$ the $J^\pi=0^-$ wave at  $E_n = 1.0$ MeV,
  obtained with the HH method as a function of $\ell_{\rm max}$ (see the
  text for details). The calculations have been performed using
  the the N3LO500 potential.}
\end{table*}
\subsection{Faddeev-Yakubovsky equations in configuration space \label{sec:FY}}

\begin{figure}[h!]
\includegraphics[width=7.5cm]{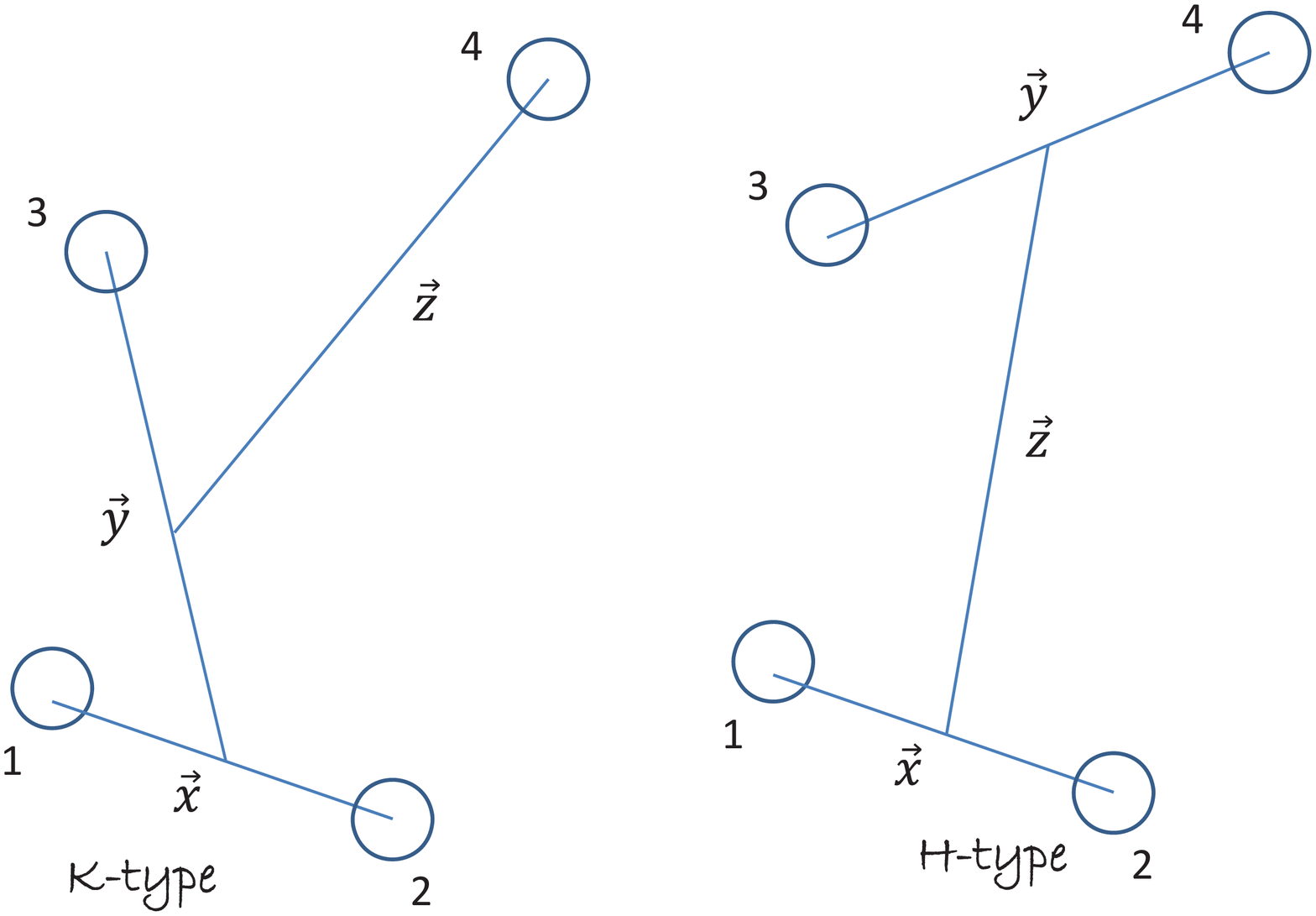}
\caption{FY's components $K$ and $H$. Asymptotically as
$z\rightarrow \infty $ components $K$ describe 3+1 particle
channels, whereas components $H$ contains asymptotic states of 2+2
channels. } \label{Fig_4b_config}
\end{figure}

In late sixties Yakubovsky~\cite{Fadd_art,Yakub} generalized  a
set of equations proposed by Faddeev~\cite{Fadd_art}, to treat
scattering problems beyond $A=3$ case. Based on the arithmetic
properties, which arise from the subsequent breaking of N-particle
system into its sub-clusters, FY formalism offers a natural way to
decompose system's wave function into the so called
Faddeev-Yakubovsky's components (FYC). As a result FYC represent
the natural structures to impose a proper wave function behavior
at the boundaries. A four-particle system requires to introduce
FYC of two distinct types: components $K$ and $H$. Asymptotes of
components $K$ incorporate 3+1
particle channels, while components $H$ contain asymptotes of 2+2
particle channels (see Fig.\ref{Fig_4b_config}). By interchanging
order of particles one can construct twelve different components
of the type $K$ and six of the type $H$. The system wave function is
then represented simply as a sum of these 18 FYC.

Employing the formalism of isospin, protons and neutrons become
two distinct states of the same particle (nucleon). For a system
of four identical particles there exist only two independent FYC,
one of type $K$ and one of type $H$. The other 16 FYC can be
obtained from the two independent FYC by applying particle
permutation operators ({\it i.e.} interchanging the order of
particles in the system). Similarly only two independent FY
equations exist; by singling out $K\equiv K_{1,23}^4$ and $H\equiv
H_{12}^{34}$, the set of two FY equations read~\cite{Rimas_03,Lea05}:
\begin{eqnarray}
\left( E-H_{0}-V_{12}\right) K &=&V_{12}(P^{+}+P^{-})\left[
(1+Q)K+H\right],
\notag \\
\left( E-H_{0}-V_{12}\right) H &=&V_{12}\tilde{P}\left[
(1+Q)K+H\right] . \label{FY1}
\end{eqnarray}%
 The particle permutation operators
$P^{+}$, $P^{-}$, $\tilde{P}$ and $Q$ represent simply different
combinations of two-particle transposition operators:
\begin{eqnarray*}
P^{+}&=&(P^{-})^{-1}=P_{23}P_{12}\ , \cr Q &=&- P_{34}\ , \cr \tilde{P}
&=&P_{13}P_{24}=P_{24}P_{13}\ .
\end{eqnarray*}
Employing the operators defined above, the wave function of the system
is given by
\begin{equation}  \label{FY_wave_func}
\Psi =\left[ 1+(1+P^{+}+P^{-})Q\right] (1+P^{+}+P^{-})K +(1+P^{+}+P^{-})(1+%
\tilde{P})H.
\end{equation}
The functions $K$ and $H$ are expanded in the basis of partial
angular momentum, spin and isospin variables, according to:
\begin{equation}
\Phi_{i}(\vec{x}_{i},\vec{y}_{i},\vec{z}_{i}) =\sum_{\alpha }
\frac{\mathcal{F}_{i}^{\alpha
}(x_{i},y_{i},z_{i})}{x_{i}y_{i}z_{i}} Y_{i}^{\alpha
}(\hat{x}_{i},\hat{y}_{i},\hat{z}_{i})\ , \label{PWB_decomp}
\end{equation}
where the functions
  $Y_{i}^{\alpha}(\hat{x}_{i},\hat{y}_{i},\hat{z}_{i})$ are defined
  below. The Jacobi coordinates, associated with each type of FYC $K$ and $H$,
are used to represent our wave functions (see
Fig.~\ref{Fig_4b_config}). Such a choice of coordinates permits
us to separate the center of mass motion and guarantees that
the kinetic energy operator is independent of angular variables. The
angular dependence is hidden in tripolar harmonics $Y_{i}^{\alpha
}(\hat{x}_{i},\hat{y}_{i},\hat{z}_{i})$, which in addition to
angular momentum variables comprise spins and isospins of the
nucleons. To couple the angular and spin quantum numbers, a
slightly different scheme is employed compared to AGS and HH methods,
namely the $jj$-coupling scheme, which is defined by: {\small
\begin{equation}
\left[ \left\{ \left( t_{i}t_{j}\right)_{t_{x}}t_{k}\right\}_{T_{3}}t_{l}%
\right]_{TT_z} \left\langle \left\{ \left[ l_{x}\left(
s_{i}s_{j}\right)_{\sigma _{x}}\right]_{j_{x}} \left[
l_{y}s_{k}\right]_{j_{y}}\right\}_{J_{3}} \left[ l_{z}s_{l}\right]
_{j_{z}} \right\rangle_{J^{\pi }J_z}
\label{K_jj_scheme}
\end{equation}%
} for the components of $K$-type, and {\small
\begin{equation}
\left[ \left( t_{i}t_{j}\right)_{t_{x}} \left( t_{k}t_{l}\right)
_{t_{y}}\right]_{TT_z} \left\langle \left\{ \left[
l_{x}\left(s_{i}s_{j}\right)_{\sigma_{x}} \right]_{j_{x}}\left[
l_{y}\left(s_{k}s_{l}\right)_{\sigma_{y}}
\right]_{j_{y}}\right\}_{j_{xy}}
l_{z}\right\rangle_{J^{\pi }J_z}
\label{H_jj_scheme}
\end{equation}%
} for the $H$-type components. Here $s_{i}=1/2$ and $t_{i}=1/2$
are the spin and isospin quantum numbers of the individual
nucleons and $\left( J^{\pi },T\right) $ are,
respectively, the total angular momentum, parity and isospin of a
four-body system. By $J_z$ and $T_z$ we identify
the projection of the system total angular momentum and isospin
on the selected axis. Nuclear Hamiltonian conserves the system
parity and angular momentum $J^{\pi}$; the system wave
function is also invariant for the rotations around the fixed axis
(so one can also fix projection quantum number $J_z$).
Furthermore the system made of two protons and two neutrons has
$T_z\equiv$0. Then each amplitude
$\mathcal{F}_{i}^{\alpha}(x_{i},y_{i},z_{i})$ is labeled by a set
of 11 non-fixed quantum numbers $\alpha$. On contrary a total
isospin $T$ of the system is not conserved, it is
allowed to take values $T=0,1$ and $2$. Combination of
different total isospin channels is necessary in order to separate
unambiguously the $\pH$ and $\nHe$ channels~\cite{Rimas_03}.

By projecting each of Eqs. (\ref{FY1}) on its natural
configuration space basis, one obtains a system of coupled
integro-differential equations. To keep the number of these equations
finite, one is obliged to introduce some additional truncations in
the partial wave expansion given in Eq.~(\ref{PWB_decomp}),
by considering only the most relevant amplitudes. This truncation
is realized by imposing the condition
$\max(l_{x},l_{y},l_{z})\leq4$ on the maximal partial angular momenta.

Equations (\ref{FY1}) are not complete as long as they are not
complemented with the appropriate boundary conditions. First, FY
amplitudes, for bound as well as for scattering states, satisfy
the regularity conditions:
\begin{equation}
\mathcal{F}_{i}^{\alpha}(0,y_{i},z_{i})=
\mathcal{F}_{i}^{\alpha}(x_{i},0,z_{i})=
\mathcal{F}_{i}^{\alpha}(x_{i},y_{i},0)=0\ .  \label{BC_xyz_0}
\end{equation}

The proper asymptotic behavior of the FY components of type-$K$ for the
scattering process is implemented in a similar way as for the HH
method, see Eq.~(\ref{eq:psica}), i.e., by splitting the FY amplitude
into two terms: a square integrable core-term
$\mathcal{F}_{C,i}^{\alpha}(x_{i},y_{i},z_{i})$ and a long-ranged
term $\mathcal{F}_{A,i}^{\alpha}(x_{i},y_{i},z_{i})$, which describes
the behavior of FY amplitudes in the far asymptotic regions,
\begin{equation}
\mathcal{F}_{i}^{\alpha}(x_{i},y_{i},z_{i})=
\mathcal{F}_{C,i}^{\alpha}(x_{i},y_{i},z_{i})+\mathcal{F}_{A,i}^{\alpha}(x_{i},y_{i},z_{i})\ .
\label{BC_xyz_inf}
\end{equation}
As explained in the previous section, the asymptotic part
$\mathcal{F}_{A,i}^{\alpha}(x_{i},y_{i},z_{i})$ is constructed
from the calculated three-nucleon wave functions (either $^3$H or
$^3$He nucleus) and involves a few parameters associated with a
scattering matrix to be determined, see Eq. $(\ref{eq:psia})$.
This term is treated as an inhomogeneous one when solving
numerically the FY equations. The core part of the 
FY partial amplitudes is expanded on the basis of Lagrange-Laguerre
mesh functions, 
employing Lagrange-mesh method~\cite{Baye_LM}:
\begin{eqnarray}
\mathcal{F}_{C,i}^{\alpha }(x_{i},y_{i},z_{i}) &=&
C^{\alpha,k,l,m} f^x_k(x_{i})f^y_l(y_{i})f^z_m(z_{i})\notag
\end{eqnarray}%
with $C^{\alpha,k,l,m}$ representing some unknown coefficients,
while the $f^x_k(x_{i})$, $f^y_l(y_{i})$ and $f^z_m(z_{i})$ are
Lagrange-Laguerre basis functions associated with each radial
variable. The set of integro-differential equations is transformed
into a linear algebra problem by projecting these equations on a
chosen three-dimensional Lagrange-Laguerre basis. The coefficients
$C^{\alpha,k,l,m}$ are determined by solving the resulting linear
algebra problem and by applying Kohn-variational principle to
determine the scattering matrix associated with the inhomogeneous
terms of Eq.~(\ref{BC_xyz_inf}).

\section{Results}
\label{sec:results}

In this section the results obtained using the three
different methods are compared between themselves as well as with
available experimental data for some selected observables. First,
in Tables  \ref{tab:nt-n3lo-1} and \ref{tab:nt-n3lo-2} we present the $\nHe$
phase-shifts and inelasticity parameters for the most relevant
waves calculated using the three different methods. Calculations
have been carried out for three different neutron laboratory
energies, $E_n=1$, $2$, and $3.5$ MeV, corresponding to cases
where experimental data exist.
In particular, we compare the parameters computed by the three
methods for the states $J^\pi=0^\pm$, $1^\pm$, and $2^-$. The
scattering in other $J^\pi$ states is dominated by the centrifugal
barrier and therefore their $S$-matrices only slightly deviates
from the unity matrix, while the results are not very sensitive to
the interaction and the method used to calculate them. Note that
the calculations of the observables has been performed including
states up to $J=4$. In all the cases, the wave function contains
states of total isospin $T=0$ and $1$.

Clearly the values of these parameters may depend on the adopted
choice of the coupling scheme between the spin of the two clusters
and the spherical harmonic function $Y_L({\bm y})$ in the
asymptotic functions $\Omega_{LS}^\pm$ (see Eq.(\ref{eq:psiom})).
As specified in the previous section, each group has chosen a
different coupling scheme. It can be shown, however, that the
(eigen)phase-shifts defined as discussed above are coupling
scheme-independent, on contrary, the mixing parameter depends on
coupling scheme (however, they are related to each other by some
constant factor). In the following we have decided to report the
mixing parameters which are proper to the coupling scheme
specified in Eq.~(\ref{eq:psiom}).

In Tables \ref{tab:nt-n3lo-1} and \ref{tab:nt-n3lo-2} we present the
inelasticity parameters, phase-shifts, and mixing parameters
for $\nHe$ scattering obtained using the N3LO500 potential
at the selected energies. By inspecting the tables, we can notice a
reasonable agreement between the results obtained by the three
different techniques.

As presented in Table \ref{tab:nt-n3lo-1}, for $0^\pm$ waves we
note a large deviation of the inelasticity
parameters from unity. In these waves, a $\nHe$ initial state will
mostly end up in a $\pH$ final state (and vice versa). The
phase-shifts are in very good agreement, only in a few cases the
results differ by more than 1\%. Larger differences are found for
the inelasticity parameters (up to 10\%), related with the
aforementioned slow convergence for these values within the HH
method. Note that the $\nHe$ $0^-$ phase-shifts are negative,
meaning that the effective interaction between the two clusters is
mostly repulsive in this wave. This is at variance with the $\pHe$
scattering case, where it was found that the interaction is
attractive for the same wave. Inelasticity parameter for $1^+$
wave is found to be close to unity. For this wave, the Pauli
repulsion keeps the incident clusters well apart, preventing
particle recombination process.
\begin{table*}[t]
\begin{ruledtabular}
\begin{tabular}{*{12}{r}}
$E_n$ & $\eta^{0+}$  & $\delta^{0+}$ & $\eta^{0-}$ & $\delta^{0-}$ &
        $\eta_{01}^{1+}$  & $\delta_{01}^{1+}$ & $\eta_{21}^{1+}$ &
        $\delta_{21}^{1+}$ & $\Re(\epsilon^{1+})$ &
  $\Im(\epsilon^{1+})$ & Method \\
(MeV) &  & (deg) & & (deg) & & (deg) & & (deg) & (deg) &(deg) & \\ \hline
1.0 & 0.267 & -63.4 & 0.553 & -10.5 & 0.998 & -31.8  & 1.000 & -0.054 & 0.306 & 0.001 & AGS \\
    & 0.267 & -63.2 & 0.545 & -10.4 & 0.997 & -31.8  & 1.000 & -0.058 & 0.228 & 0.001 & FY\\
    & 0.259 & -64.0 & 0.505 & -10.5 & 0.998 & -32.4  & 1.000 & -0.065 & 0.297 & 0.001 & HH\\
\hline
2.0 & 0.162 & -81.4 & 0.380 & -12.9 & 0.997 & -43.8 & 1.000 & -0.220 &0.585 & 0.003 & AGS \\
    & 0.162 & -81.4 & 0.368 & -12.7 & 0.997 & -43.8 & 1.000 & -0.184 &0.467 & 0.006 & FY \\
    & 0.147 & -82.2 & 0.348 & -12.9 & 0.996 & -44.3 & 1.000 & -0.250 &0.574 & 0.002 & HH \\
\hline
3.5 & 0.086 & -97.6 & 0.093 & -8.48 & 0.996 & -56.0 & 1.000 & -0.567 &0.971 & 0.007 & AGS \\
    & 0.086 & -98.2 & 0.075 & -8.41 & 0.996 & -56.3 & 1.000 & -0.566 &0.884 & 0.003 & FY \\
    & 0.074 & -98.5 & 0.088 & -8.16 & 0.996 & -56.2 & 1.000 & -0.618 &0.957 & 0.004 & HH \\
\end{tabular}
\end{ruledtabular}
\caption{ \label{tab:nt-n3lo-1}
$\nHe$ inelasticity parameters  $\eta_{LS}^{J\pi}$, phase-shifts $\delta_{LS}^{J\pi}$, and mixing parameters
 $\epsilon^{J\pi}$ for the $J^\pi=0^\pm$ and $1^+$ waves
  at  $E_n = 1.0$, 2.0, and 3.5 MeV, obtained with the three methods
  described in the text. The phase-shifts and mixing parameters are
  given in degrees. The calculations have been performed using
  the N3LO500 potential.
}
\end{table*}

Let us now inspect Table~\ref{tab:nt-n3lo-2}. For the $1^-$ state,
it is possible to note that the $LS=10$ (${}^1P_1$) phase-shift is
negative showing that the interaction of the $\nHe$ clusters is
repulsive (again, for this wave the $\pHe$ phase-shift is
positive). In this case, the inelasticity parameter deviates
sizably from unity. On the other hand, the $LS=11$ (${}^3P_1$)
phase-shift is positive and large as for $\pHe$, while
$\eta_{11}^{1-}\approx 1$. We note a good agreement between the
results obtained by the three different methods for these
parameters, and also for the mixing parameter $\epsilon^{1-}$. For
the $2^-$ state, the $LS=11$ (${}^3P_2$) phase-shift is positive
and large as for $\pHe$, while the corresponding inelasticity
parameter decreases at low energies and then reaches a sort of
plateau between $E_n=2$ and $3.5$ MeV. There is a good agreement
for the phase-shifts, while for the inelasticity parameters we
observe again a somewhat sizable deviation between the results 
obtained with the HH and the AGS/FY methods, again connected to the slow convergence
of the HH expansion. The $LS=31$ (${}^3F_2$) phase-shift
$\delta_{31}^{2-}$ and mixing parameter $\epsilon^{2-}$ are rather
small, due to the large centrifugal barrier (in this case
$\eta_{31}^{2-}$ is very close to $1$). In any case, we observe a
reasonable agreement between different calculation methods even
for these tiny quantities.

\begin{table*}[t]
\begin{ruledtabular}
\begin{tabular}{*{14}{r}}
$E_n$ & $\eta_{10}^{1-}$  & $\delta_{10}^{1-}$ & $\eta_{11}^{1-}$ &
        $\delta_{11}^{0-}$ & $\Re(\epsilon^{1-})$ &
        $\Im(\epsilon^{1-})$ &
        $\eta_{11}^{2-}$  & $\delta_{11}^{2-}$ & $\eta_{31}^{2-}$ &
        $\delta_{31}^{2-}$ & $\Re(\epsilon^{2-})$ &
        $\Im(\epsilon^{2-})$ & Method \\
(MeV) &  & (deg) & & (deg) & (deg) & (deg) & & (deg) & & (deg) & (deg) &(deg) & \\ \hline
1.0 & 0.959 & -0.263 & 0.994 & 7.43 & -1.98 & -0.869 & 0.923 & 17.5 &1.000 & 0.003 & -0.249 & -0.093 & AGS \\
    & 0.958 & -0.333 & 0.995 & 7.57 & -2.02 & -0.849 & 0.938 & 17.3 &1.000 & 0.002 & -0.208 & -0.092 & FY\\
    & 0.957 & -0.295 & 0.993 & 7.57 & -2.02 & -0.909 & 0.938 & 17.0 &1.000 & 0.003 & -0.248 & -0.086 & HH\\
\hline
2.0 & 0.864 & -0.806 & 0.985 & 20.0 & -1.96 & -1.32 & 0.665 & 47.1 &1.000 & 0.021 & -0.330 & -0.286 & AGS \\
    & 0.862 & -0.750 & 0.986 & 20.2 & -1.96 & -1.36 & 0.685 & 47.0 &1.000 & 0.012 & -0.301 & -0.248 & FY \\
    & 0.859 & -0.865 & 0.989 & 20.5 & -1.94 & -1.37 & 0.715 & 47.0 &1.000 & 0.022 & -0.334 & -0.283 & HH \\
\hline
3.5 & 0.699 & -2.60 & 0.992 & 38.6 & -1.90 & -1.65 & 0.676 & 70.8 &1.000 & 0.101 & -0.237 & -0.422 & AGS \\
     & 0.694 & -2.65 & 0.990 & 38.1 & -1.87 & -1.75 & 0.681 & 70.4 &1.000 & 0.078 &  -0.255  & -0.415 &FY \\
    & 0.694 & -2.56 & 0.994 & 37.3 & -1.87 & -1.75 & 0.714 & 69.1 &1.000 & 0.092 & -0.259 & -0.408 & HH \\
\end{tabular}
\end{ruledtabular}
\caption{ \label{tab:nt-n3lo-2}
Same as for Table but for
the $J^\pi=1^-$ and $2^-$ waves.}
\end{table*}

Let us now examine how the good agreement found for the
phase-shifts and mixing parameters calculated with the three
different methods is reflected in the experimental observables.
Let us start with a case of $\nHe$ elastic observables. We have
considered the differential cross section, the neutron analyzing
power $A_{y0}$, the  $\He$  analyzing power $A_{0y}$, and the spin
correlation coefficient $A_{yy}$. The analyzing power observables
are rather sensitive to small variations of the P-wave
phase-shifts in the kinematical regime considered in this paper,
while $A_{yy}$ is  also sensitive to the S-wave phase shifts.

\begin{figure}[hbt]
\begin{center}
\includegraphics[scale=0.5,clip]{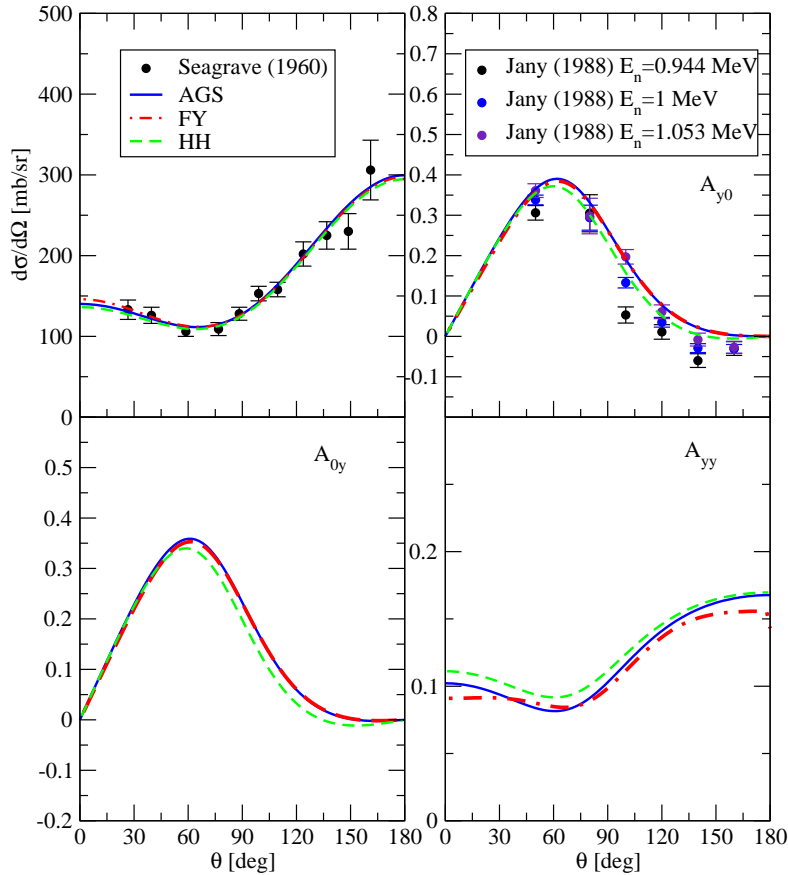}
\end{center} 
\caption{ \label{fig:nhe1} (Color online) Differential cross
section, proton analyzing power $A_{y0}$, $\He$ analyzing power
$A_{0y}$, and spin correlation coefficient $A_{yy}$ for $\nHe$
elastic scattering at  $E_n=1$ MeV neutron lab energy obtained using
the N3LO500 potential. The lines show the results
obtained using the AGS (blue solid lines), FY (red dot-dash lines),
and the HH (green dashed lines) methods. The experimental data are from
Refs.~\cite{Sea60,Jany88}.
}
\end{figure}

\begin{figure}[hbt]
\begin{center}
\includegraphics[scale=0.5,clip]{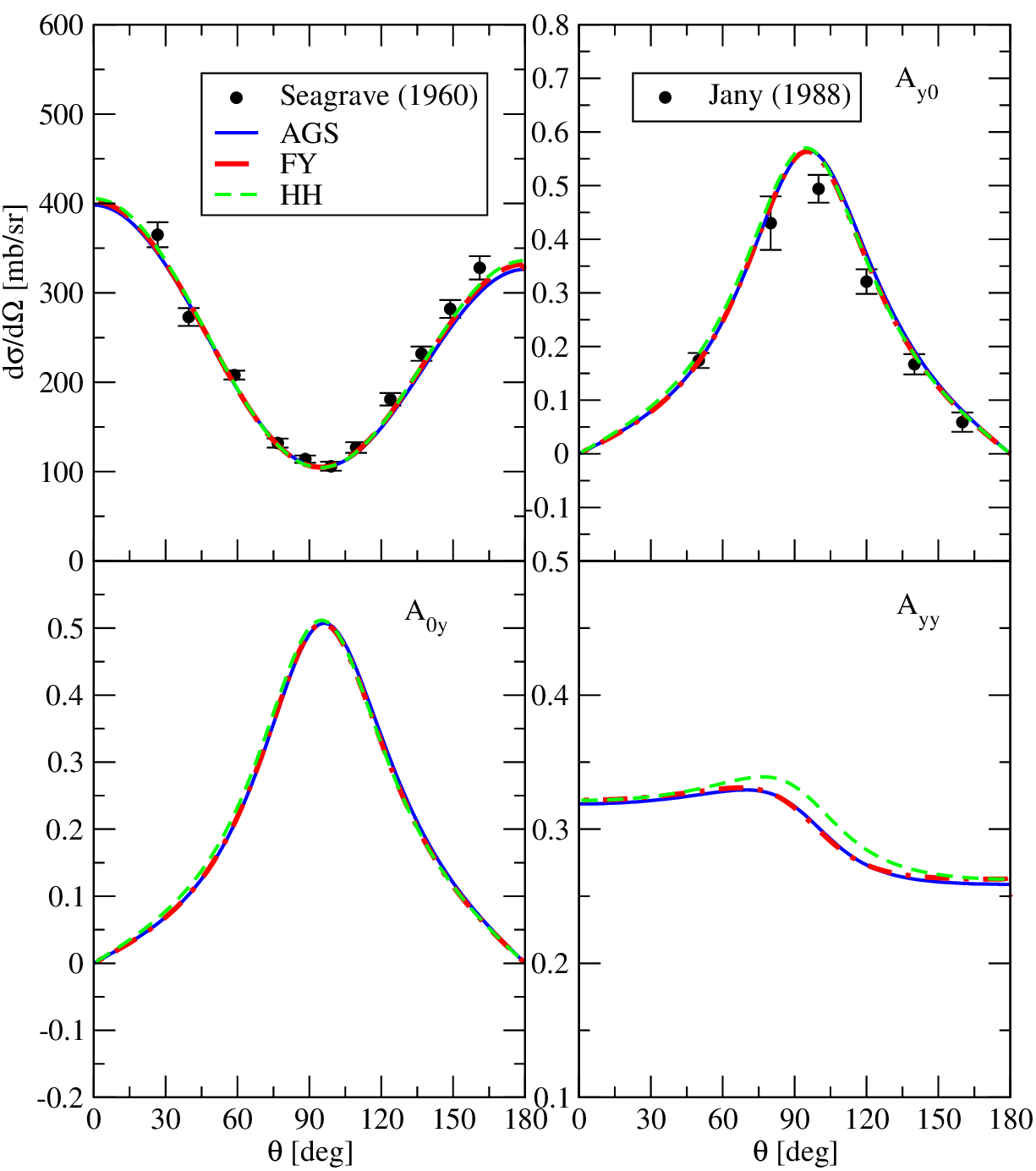}
\end{center} 
\caption{ \label{fig:nhe2} (Color online) Same as
  Fig.~\protect\ref{fig:nhe1} but for neutron energy $E_n=2$
  MeV.  The experimental data are from
Refs.~\cite{Sea60,Jany88}.
}
\end{figure}

\begin{figure}[hbt]
\begin{center}
\includegraphics[scale=0.5,clip]{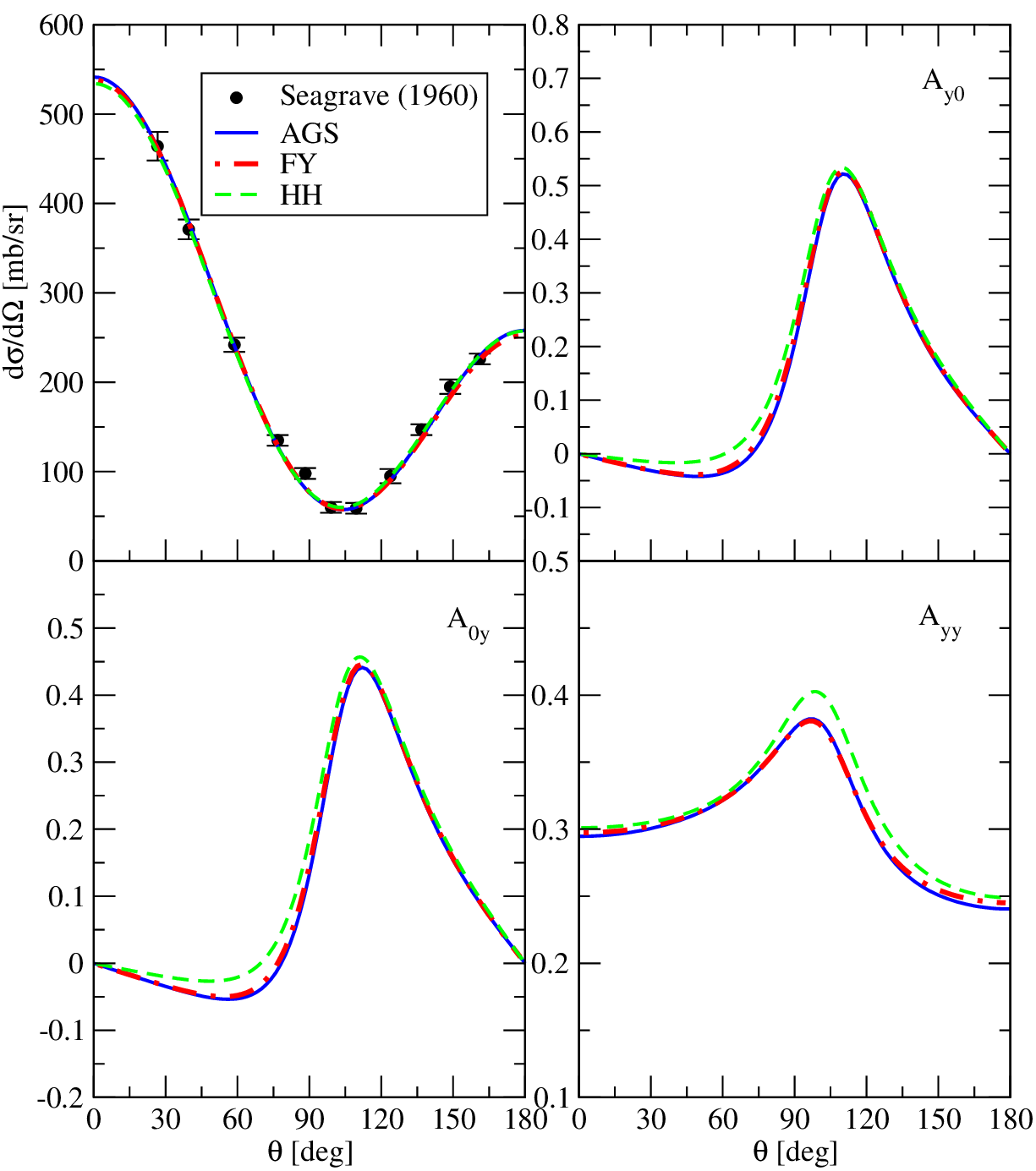}
\end{center} 
\caption{ \label{fig:nhe3} (Color online) Same as
  Fig.~\protect\ref{fig:nhe1} but for neutron energy $E_n=3.5$
  MeV.  The experimental data are from
Ref.~\cite{Sea60}.
}
\end{figure}

In Figs.~\ref{fig:nhe1}, \ref{fig:nhe2}, and \ref{fig:nhe3} we
have reported the results obtained using the AGS equation (blue
solid lines), the HH expansion method (green dashed lines), and
the FY equations (red dot-dash lines) using the N3LO500 potential
for $E_n=1$, $2$, and $3.5$ MeV, respectively.  Where available,
we compare the calculated observables with the experimental data.
As can be seen by inspecting the three figures, for the
differential cross section the three curves almost always
perfectly coincide and can be hardly distinguished. For the
$A_{y0}$ and $A_{0y}$ analyzing power observables, the AGS and FY
results almost coincide, while the HH results slightly differ
(however, we note that the differences between the three
calculations are in any case within the experimental
error-bars). We also note in Fig.~\ref{fig:nhe1} a rather strong
energy-dependence of the analyzing power from the
measurements at $E_n=0.944$, $1$, and $1.053$ MeV~\cite{Jany88}.
For the $A_{yy}$ observable, the predictions obtained by the three
methods are slightly at variance. This observable, as already
stated, is quite sensitive to the $0^+$ phase shift, for which the
convergence of the three methods is more problematic.  In spite of
these difficulties, the agreement in the considered observable is
still acceptable.

\begin{figure}[hbt]
\begin{center}
\includegraphics[scale=0.5,clip]{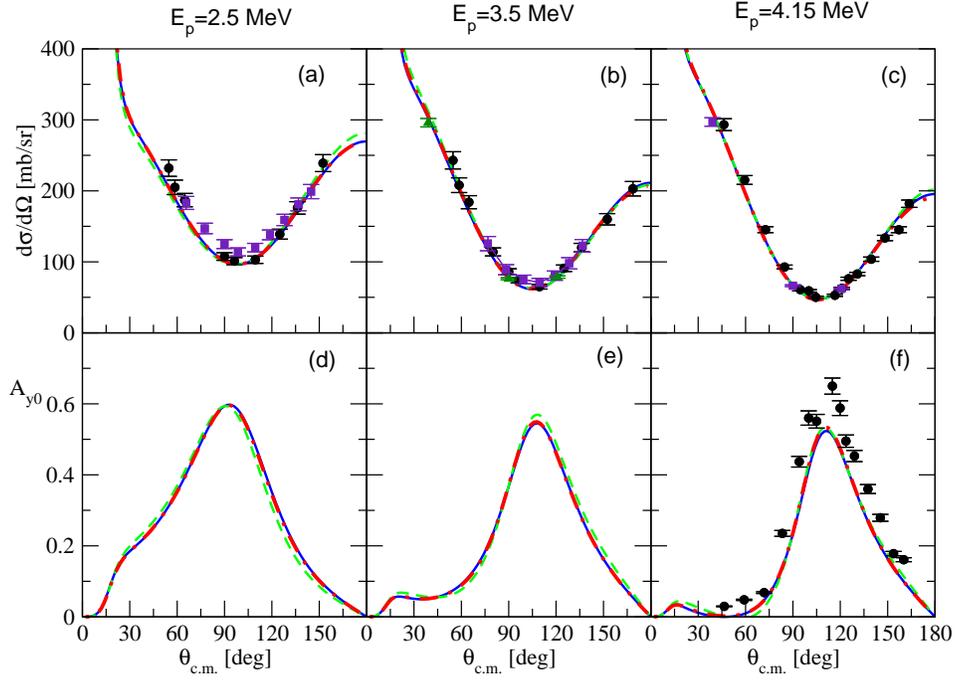}
\end{center} 
\caption{ \label{fig:pt} (Color online) Differential cross
sections (upper panels) and proton analyzing powers $A_{y0}$ (lower panels) for $\pH$
elastic scattering at  $E_p=2.5$, $3.5$, and $4.15$ MeV proton 
energies obtained using the N3LO500 potential.
The lines show the results
obtained using the AGS (blue solid lines), FY (red dot-dash lines),
and the HH (green dashed lines) methods. In many cases, the curves
overlap and cannot be distinguished. The experimental data in
panel (a) are from Refs.~\cite{Clas51} (circles) and~\cite{Hemme49} (squares),
in panel (b) from Refs.~\cite{Clas51} (circles),~\cite{Mandu68}
(squares), and~\cite{Iva68} (triangles), in panel (c) from
Refs.~\cite{Kanko76} (circles) and~\cite{Iva68} (squares), and finally
in panel (f) from Ref.~\cite{Kanko76} (circles). 
}
\end{figure}

Let us now consider the $\pH$ elastic observables. For this case, we
have decided to show the comparison of the theoretical results
for the differential cross section and the proton analyzing power
$A_{y0}$, where experimental data are available. These observables are
reported in Fig.~\ref{fig:pt} for three different energies of the
proton beam, $E_p=2.5$, $3.5$ and $4.15$ MeV. In the upper panels, we
have reported the differential cross sections and in the lower
panels the proton analyzing power $A_{y0}$. As it can be seen, the AGS
and FY results are almost indistinguishable for all the considered
cases. The HH results show somewhere a slight deviation from the
AGS/FY values, probably due to the slow convergence observed for
the inelasticity parameters.

Regarding the comparison with the experimental data, we note again a
good reproduction of the differential cross sections at all the
energies. For $A_{y0}$ at $E_p=4.15$ MeV, the only case for which
there are experimental data for
this observable, we note a slight underprediction of the peak and at
forward angles, where the nuclear and Coulomb scattering amplitudes
interfere.

\begin{figure}[hbt]
\begin{center}
\includegraphics[scale=0.5,clip]{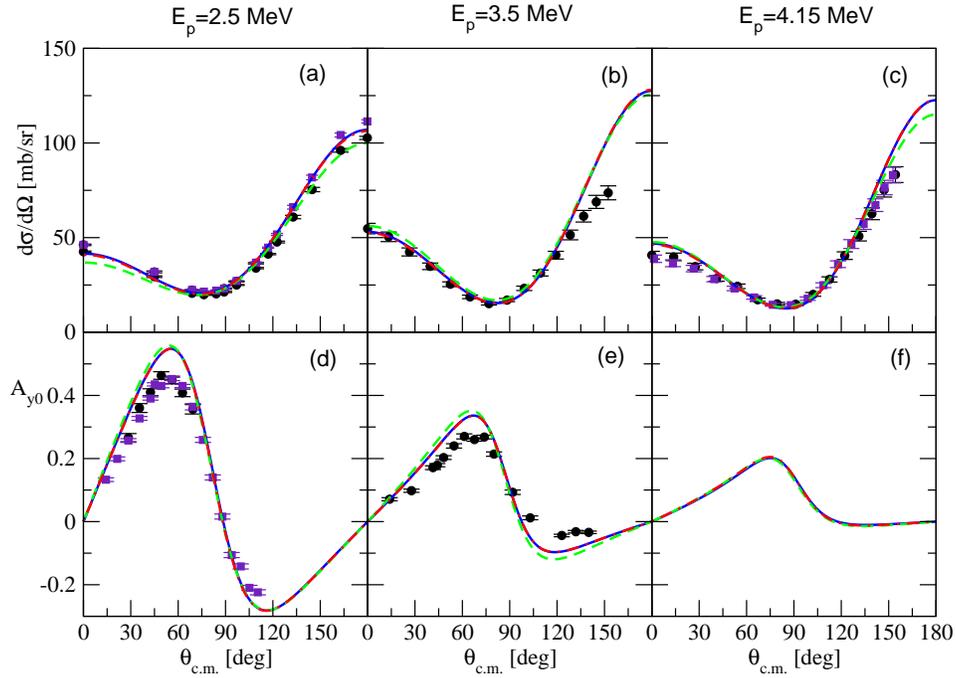}
\end{center} 
\caption{ \label{fig:ptx} (Color online) Same as Fig.~\protect\ref{fig:pt}
but for $\pH\rightarrow \nHe$ process. The experimental data in
panel (a) are from Ref.~\cite{Drosg80} (circles),
in panel (b) from Ref.~\cite{Jarvis56} (circles), in panel (c) from
Refs.~\cite{Jarvis56} (circles) and~\cite{Will53} (squares), in panel
(d) from Refs.~\cite{Tornow81} (circles) and~\cite{Doy81} (squares),
and finally in panel (e) from Ref.~\cite{Doy81} (circles).
}
\end{figure}

In Fig.~\ref{fig:ptx}, the comparison is extended to the reaction
$\pH\rightarrow \nHe$. Also in this case, we have reported the
differential cross section and proton analyzing power at $E_p=2.5$,
$3.5$ and $4.15$ MeV. The $\pH\rightarrow \nHe$
differential cross section has been found to depend strongly on the
$2^-$ wave, and it is therefore rather sensitive to the convergence of
the calculations. For this reason, we observe more sizeable
deviations between the AGS/FY and HH results. In any case, from panels
(b) and (c), it is possible to note a clear discrepancy among the
theoretical calculations and the experimental data at backward
angles. Regarding $A_{y0}$, we observe a fair agreement between the
theoretical results. Here, we note an overprediction of the
calculated $A_{y0}$ in the maximum region with respect to the
available experimental data. At $E_p=3.5$ MeV, see panel (e), theoretical
calculations and data also disagree in the region of the minimum. The
origin of these discrepancies  is still not clear. 
Moreover, we note that in Ref.~\cite{DF07c}, using four realistic
NN potentials, a significant sensitivity of charge-exchange
observables to the NN force model has been found.

\section{Conclusions}
\label{sec:conc}

In this work, we have studied some low energy $\nHe$ and $\pH$
elastic and charge-exchange observables by using three different
approaches, the HH, AGS and FY techniques. Very accurate solutions
of the 4N scattering problem using the AGS
technique~\cite{DF07,DF07b,DF07c} were obtained already a few
years ago. The long-range Coulomb interaction in this approach is
taken into account using the screening  and renormalization
method~\cite{Alt78,DFS05}.  In recent years, after adding some
additional numerical power, also the accuracy of the calculations
performed using the HH and FY techniques
increased~\cite{Vea09,Vea13,Lazaus09}. Therefore, it becomes quite
interesting to compare the results obtained by the different
methods in order to test their capability to solve the 4N
scattering problem. Around five years ago, some of the authors of
the present paper presented a very detailed comparison for $\pHe$
and $\nH$ observables~\cite{Deltuva11}. The aim of the present
paper is to extend the benchmark to the $\nHe$ and $\pH$
scattering.

Here we have shown that for N3LO500 potential  the
results obtained by the different techniques are in good
agreement. In particular, FY and AGS results are in a very good
agreement. The phase-shifts, mixing angles and
observables calculated using the HH method show some small
deviations from those obtained by AGS/FY techniques. Anyway, the
differences are tiny, and usually do not exceed the experimental
errors. Therefore, we can conclude that all the considered
theoretical methods have reached a rather high level of accuracy
in the description of $\nHe$ and $\pH$ elastic and charge-exchange
scattering making comparison with experiment reliable and
meaningful.

Concerning the comparison with the experiments, in most of the
cases we have observed a good agreement between the results
obtained using the N3LO500 potential and the available
experimental data. Some disagreements persist for the analyzing
power, and for the $\pH\rightarrow \nHe$ differential cross
section at backward angles. These observables show also a sizable
 NN interaction model dependence~\cite{DF07c}. Therefore
 as a paramount test of nuclear interaction models, it will be
 interesting to explore the effect of the inclusion of a 3N
interaction. Work in this direction is in progress~\cite{Vea16}.

\bigskip

\bibliographystyle{prsty}

\end{document}